\documentclass[a4paper,11pt]{article}
\usepackage{jcappub} 
\usepackage{lineno}

\usepackage{comment}
\usepackage{graphicx}
\usepackage{subfigure}
\usepackage{palatino}
\usepackage[commandnameprefix=always]{changes}
\usepackage{hyperref}
\hypersetup{colorlinks=true,linkcolor=blue,urlcolor=blue,citecolor=blue}
\usepackage[toc,page]{appendix}
\usepackage[normalem]{ulem}

\usepackage{orcidlink}
\usepackage{lipsum}
\usepackage{graphicx}
\usepackage{subfigure}
\usepackage{palatino}
\usepackage{sans}
\usepackage{adjustbox}
\usepackage{latexsym}
\usepackage{amsmath}
\usepackage{amssymb}
\usepackage{amsfonts}
\usepackage{dcolumn}
\usepackage{bm}
\usepackage{tikz}
\usepackage{bigints}
\usepackage{array,tabularx,multirow,booktabs}
\usepackage[tracking=true]{microtype}
\SetTracking{}{500}
\SetTracking{encoding={*}, shape=sc}{40}
\UseRawInputEncoding 
\allowdisplaybreaks
\usepackage{adjustbox}
\usepackage{latexsym}
\usepackage{amsmath}
\usepackage{amssymb}
\usepackage{amsfonts}
\usepackage{dcolumn}
\usepackage{bm}
\usepackage{tikz}
\usepackage{bigints}
\usepackage{array,tabularx,multirow,booktabs}
\usepackage[tracking=true]{microtype}
\usepackage{color}

\def\1o2{{1\over2}}

\UseRawInputEncoding 
\allowdisplaybreaks

\arxivnumber{2504.07561} 
\title{\boldmath Regular Black Hole Models in the Transition from Baryonic Matter to Quark Matter}

\author[a,b,c]{V. Vertogradov
\orcidlink{0000-0002-5096-7696}}
\affiliation[a]{Physics department, Herzen state Pedagogical University of Russia,
48 Moika Emb., Saint Petersburg 191186, Russia.}
\affiliation[b]{Center for Theoretical Physics, Khazar University, 41 Mehseti Street, Baku, AZ-1096, Azerbaijan.}
\affiliation[c]{SPB branch of SAO RAS, 65 Pulkovskoe Rd, Saint Petersburg
196140, Russia.}

\author[d]{A. \"Ovg\"un
\orcidlink{0000-0002-9889-342X}
}
\emailAdd{ali.ovgun@emu.edu.tr}
\affiliation[d]{Physics Department, Eastern Mediterranean
University, Famagusta, 99628 North Cyprus, via Mersin 10, Turkiye.}

\abstract{In this paper, we investigate gravitational collapse scenarios involving baryonic matter transitioning into quark-gluon plasma under extreme astrophysical conditions, focusing on their implications for the formation of regular black holes. Standard gravitational collapse models inevitably predict central singularities, highlighting the limitations of classical general relativity in extreme density regimes. By introducing a physically motivated, inhomogeneous transition rate between baryonic and quark matter, we demonstrate analytically and numerically that it is possible to construct regular black hole solutions featuring a nonsingular de Sitter-like core. We further analyze the observable consequences of these models, particularly emphasizing modifications to the black hole shadow radius, which provide direct observational constraints accessible through Event Horizon Telescope (EHT) measurements.}

\begin{document}
\maketitle
\flushbottom

\section{Introduction}

After the groundbreaking observations by the Event Horizon Telescope (EHT) collaboration, which captured images of black holes at the centers of both the M87 galaxy and our own Milky Way, black holes have emerged as central objects of study in contemporary theoretical and observational physics \cite{EventHorizonTelescope:2019dse,EventHorizonTelescope:2022wkp,Vagnozzi:2022moj}. Despite the remarkable success of General Relativity (GR) in describing gravitational phenomena, the theory exhibits significant limitations when applied to regions of extreme spacetime curvature, such as black hole singularities \cite{Bambi:2023try,Ansoldi:2008jw,Lan:2023cvz}. Classical black hole solutions, like the Schwarzschild spacetime, predict singularities where curvature invariants diverge, indicating the breakdown of classical GR and suggesting the necessity of a more comprehensive theoretical framework. For example, the Schwarzschild black hole solution contains two distinct singular features: a coordinate singularity at the event horizon (\(r = 2M\)) and a physical singularity at the center (\(r = 0\)). While the event horizon singularity arises solely due to the choice of coordinates and can be removed via a suitable coordinate transformation (such as transitioning to Kruskal-Szekeres coordinates), the central singularity is a genuine curvature singularity where spacetime invariants diverge and cannot be eliminated by any change of coordinates.

The seminal singularity theorem by Roger Penrose rigorously demonstrated that gravitational collapse under physically reasonable conditions (particularly the strong energy condition) inevitably leads to singularity formation \cite{Penrose:1964wq,Penrose:1969pc}. In contrast, Bardeen introduced a notable exception a regular black hole solution by explicitly violating the strong energy condition through the introduction of exotic matter \cite{Bardeen1968qtr..conf...87B}. This class of singularity-free solutions, known collectively as \textit{regular black holes}, is characterized by the presence of a nonsingular central region commonly referred to as a \textit{de Sitter core} \cite{Dymnikova:2015yma}. As initially suggested by Sakharov and Gliner, the emergence of this core can be physically interpreted as a phase transition occurring at extremely high densities, causing baryonic matter to transition into a vacuum-like state described effectively by the de Sitter metric \cite{Sakharov1966JETP...22..241S,Gliner1966JETP...22..378G}. \textcolor{black}{It is important to note that a de Sitter core is not the only possible structure at the center of a regular black hole. Simpson and Visser demonstrated that a regular center can also be formed using a Minkowski core~\cite{Simpson:2018tsi}. They introduced the black-bounce metric, a singularity-free spacetime that smoothly interpolates between a regular black hole and a traversable wormhole \cite{Simpson:2018tsi}.
 Simpson, Martin-Moruno, and Visser extend the black-bounce framework to Vaidya spacetimes, demonstrating how radiating geometries can dynamically interpolate between black-bounce black holes and traversable wormholes \cite{Simpson:2019cer}. Franzin et al. introduce a novel family of charged black-bounce spacetimes that smoothly interpolate between regular black holes and traversable wormholes under the influence of electromagnetic charge \cite{Franzin:2021vnj}. Lobo et al. construct a broad class of novel black-bounce spacetimes that unify regular black holes and traversable wormholes while thoroughly examining their regularity, energy condition violations, and causal structure \cite{Lobo:2020ffi}. Borissova, Liberati, and Visser analyze how kinematical transitions among singular and regular black holes, horizonless compact objects, and cosmological bounces inherently involve violations of the null convergence condition \cite{Borissova:2025msp}.
 Carballo-Rubio et al. propose a framework for singularity-free gravitational collapse that transitions smoothly from regular black holes to horizonless compact objects without encountering curvature singularities \cite{Carballo-Rubio:2023mvr}.
 Mazza and Liberati investigate the existence and properties of regular black holes and horizonless ultra-compact objects arising in Lorentz-violating gravity frameworks, elucidating their spacetime structure and stability conditions \cite{Mazza:2023iwv}.
 Carballo-Rubio et al. critically assess the proposal of regular evaporating black holes with stable cores, highlighting key misconceptions and reinforcing the physical constraints necessary for their consistency \cite{Carballo-Rubio:2022pzu}.
 Carballo-Rubio et al. uncover a theoretical connection between regular black hole geometries and horizonless ultracompact star solutions, showing how both can be described within a unified metric framework without singularities or event horizons \cite{Carballo-Rubio:2022nuj}.
 Franzin, Liberati, Mazza, and Vellucci construct a class of stable rotating regular black hole solutions by generalizing non-singular metrics to include angular momentum and analyzing their perturbative stability \cite{Franzin:2022wai}.
 Carballo-Rubio et al. propose a family of regular black hole models in which modifications to the interior geometry prevent the usual mass inflation instability at the inner horizon \cite{Carballo-Rubio:2022kad}.
 Carballo-Rubio et al. investigate the instability of the inner horizon in regular black hole spacetimes and demonstrate how perturbations can lead to the development of unstable cores even in non-singular geometries \cite{Carballo-Rubio:2021bpr}.
 Carballo-Rubio et al. critically evaluate the theoretical viability of regular black hole models by examining the geometric and physical conditions—such as energy requirements and asymptotic behavior—necessary to avoid singularities while remaining consistent with general relativity \cite{Carballo-Rubio:2018pmi}.
 Casadio, Kamenshchik, and Ovalle develop regularized Schwarzschild black hole solutions and investigate their integration into cosmological models, highlighting the impact of non-singular geometries on early-universe dynamics \cite{Casadio:2025pun}.
 Ovalle introduces a gravitational decoupling method that systematically transforms perfect fluid solutions into anisotropic fluid configurations, broadening the repertoire of physically viable compact object models in general relativity \cite{Ovalle:2017fgl}.
 Ovalle, Casadio, da Rocha, and Sotomayor extend the gravitational decoupling approach to generate a wide class of anisotropic fluid solutions from known isotropic metrics, illustrating the method’s versatility in modeling compact objects \cite{Ovalle:2017wqi}.
 Gabbanelli, Ovalle, Sotomayor, Stuchlík, and Casadio construct a causal interior solution for a Schwarzschild–de Sitter spacetime using the gravitational decoupling method, ensuring regularity and physical viability at the matching surface \cite{Gabbanelli:2019txr}. Ovalle, Casadio, and Giusti introduce a Minkowski deformation technique to construct regular hairy black hole solutions featuring nontrivial scalar hair without central singularities \cite{Ovalle:2023ref}.}

\textcolor{black}{Among the first and most influential solutions in this context is the Bardeen regular black hole, initially proposed by Bardeen in 1968 \cite{Bardeen1968qtr..conf...87B}. The Bardeen solution is characterized by a central region where curvature invariants remain finite and regular due to the presence of an effective magnetic monopole source.   Although initially introduced as a purely phenomenological model, it was later interpreted physically by Ayon-Beato and Garcia as an exact solution arising from general relativity coupled to nonlinear electrodynamics (NED) \cite{Ayon-Beato:1998hmi,Ayon-Beato:1999kuh}. Ayon-Beato and Garcia construct a singularity-free black hole solution within general relativity by coupling to a suitably chosen nonlinear electrodynamics Lagrangian that regularizes the central core \cite{Ayon-Beato:1998hmi}. Then Ayon-Beato and Garcia derive a new singularity-free black hole solution within general relativity by coupling to a tailored nonlinear electrodynamics Lagrangian that ensures regularity at the core and the correct asymptotic behavior \cite{Ayon-Beato:1999kuh}. Ayon-Beato and Garcia present a nonsingular charged black hole solution sourced by nonlinear electrodynamics, demonstrating regular behavior at the core while maintaining appropriate asymptotic limits \cite{Ayon-Beato:1999qin}. Ayon-Beato and Garcia reinterpret the Bardeen regular black hole as a nonlinear magnetic monopole solution in general relativity, demonstrating how a suitable nonlinear electrodynamics source reproduces the Bardeen metric without singularities \cite{Ayon-Beato:2000mjt}. Subsequently, Hayward introduced another seminal regular black holes model in 2006, explicitly constructed to represent a physically plausible black hole formation and evaporation scenario \cite{Hayward:2005gi}. The Hayward black hole model achieves regularity by employing a mass function carefully designed to avoid singularities, leading to an interior that resembles a de Sitter-like vacuum. Beyond simply providing a regular interior geometry, Hayward’s solution offers important insights into black hole evaporation and Hawking radiation processes, serving as a theoretical prototype for understanding the end stages of black hole evolution within a regular framework. Together, the Bardeen and Hayward solutions have significantly influenced the field by highlighting the viability of singularity-free spacetimes. They have become reference benchmarks for testing new physical theories and modified gravity models, contributing greatly to our understanding of fundamental gravitational physics.}

 {\color{black}Recent developments have considerably revived interest in regular black hole solutions, motivated by both theoretical elegance and observational prospects. Various models have been proposed, predominantly relying on exotic sources such as nonlinear electrodynamics. 
 Bronnikov constructs regular magnetic black hole and monopole solutions within general relativity by employing a nonlinear electrodynamics source that smooths out the central singularity \cite{Bronnikov:2000vy}. Bronnikov proposes a class of regular black hole solutions that serve as an alternative to black-bounce geometries, detailing their metric properties and physical viability \cite{Bronnikov:2024izh}. Bronnikov investigates how nonlinear electrodynamics can give rise to regular black hole and wormhole solutions by identifying suitable Lagrangian functions that eliminate central singularities \cite{Bronnikov:2017sgg}. Bronnikov, Konoplya, and Zhidenko analyze the linear stability of wormholes and regular black hole solutions supported by a phantom scalar field and show that such configurations exhibit dynamical instabilities under perturbations \cite{Bronnikov:2012ch}. Bronnikov, Melnikov, and Dehnen introduce the concept of black universes, presenting regular black hole solutions that connect to expanding cosmological regions beyond their horizons without encountering singularities \cite{Bronnikov:2006fu}. Bolokhov, Bronnikov, and Skvortsova present a class of spacetimes featuring a regular central core in place of a black-bounce throat, demonstrating how specific matter configurations can yield nonsingular geometries without bounce behavior \cite{Bolokhov:2024sdy}. However, these constructions typically rely on theoretically problematic assumptions, such as the existence of magnetic monopoles at the core. Fan and Wang develop a general procedure for constructing regular black hole solutions in general relativity by coupling to nonlinear electrodynamics, ensuring nonsingular metrics that respect energy conditions \cite{Fan:2016hvf}. Fan investigates the critical thermodynamic phenomena of regular black holes in anti-de Sitter spacetimes, revealing phase transition behavior analogous to Van der Waals fluids \cite{Fan:2016rih}. Toshmatov, Ahmedov, Abdujabbarov, and Stuchlik derive a rotating regular black hole solution by applying the Newman-Janis algorithm to a non-singular static metric coupled to nonlinear electrodynamics, ensuring the absence of curvature singularities on and off the axis \cite{Toshmatov:2014nya}. Toshmatov, Stuchlik, and Ahmedov construct a generic class of rotating regular black holes in general relativity coupled to nonlinear electrodynamics, characterizing their horizon structure and energy condition satisfaction for arbitrary rotation parameters \cite{Toshmatov:2017zpr}. Malafarina and Toshmatov demonstrate a deep connection between regular black hole solutions in nonlinear electrodynamics and semiclassical dust collapse models, showing how effective matter profiles in collapse can reproduce known nonsingular geometries \cite{Malafarina:2022oka}.}

It is widely accepted that strange quark matter, comprising up, down, and strange quarks, represents the most stable configuration of baryonic matter from an energetic viewpoint. Witten \cite{Witten:1984rs} has proposed two primary scenarios for the formation of strange matter: one involving the quark-hadron transition occurring in the primordial universe, and another through the transformation of neutron stars into strange quark stars under conditions of extremely high density. Certain strong interaction theories, particularly quark bag models, hypothesize a phase of vacuum symmetry breaking within hadrons. This leads to significant differences between vacuum energy densities inside and outside hadrons, resulting in a vacuum-induced pressure on the quark confinement boundary (bag wall) \cite{Farhi:1984qu}, which balances internal quark pressures and stabilizes the hadronic structure.

Various mechanisms have been suggested to explain the formation of quark stars. A primary scenario involves core collapse within massive stars after supernova explosions, triggering first- or second-order phase transitions into deconfined quark matter \cite{Cheng:1998na}. Proto-neutron stars or existing neutron star cores provide particularly favorable conditions for the conversion of standard nuclear matter into stable strange quark matter \cite{Dai:1995uj}. Additionally, neutron stars in low-mass X-ray binaries may accumulate enough mass through accretion to induce a phase transition to strange quark matter, resulting in the formation of strange stars \cite{Cheng:1995am}. Recent studies have explored the structure and properties of neutron stars and strange quark stars within both standard and modified theories of gravity, incorporating effects such as anisotropy, non-linear equations of state, and dark matter condensation \cite{Ozel:2016oaf,Olmo:2019flu,Deb:2016lvi,Rahaman:2014waa,Panotopoulos:2017eig,Lopes:2019psm,Panotopoulos:2017pgv,Panotopoulos:2019zxv,Panotopoulos:2020zqa,Panotopoulos:2019wsy,Panotopoulos:2017jdc,Panotopoulos:2020uvq,Panotopoulos:2024jtn}. Consequently, investigating the collapse of strange quark matter is crucial both for enhancing our understanding of the dynamics and evolution of strange quark stars and addressing fundamental issues within the framework of general relativity.

A novel exact solution is introduced, extending earlier models of gravitational collapse such as those by Vaidya \cite{Vaidya:1951zz}, Bonnor and Vaidya \cite{Bonnor:1970zz}, Lake and Zannias \cite{Lake:1991bff}, and Husain \cite{Husain:1995bf} to scenarios involving strange matter. For a broader framework on deriving spherically symmetric configurations within the Vaidya geometry, one may consult Ref. \cite{Wang:1998qx}. Additionally, extensive discussions on the behavior and characteristics of singularities emerging during collapse in Vaidya-type spacetimes are available in Refs. \cite{Petrov:2023otl,Mkenyeleye:2014dwa,Ghosh:2008jca,Hossenfelder:2009fc,Nasereldin:2023qph,Ghosh:2000ud,Ghosh:2000bc,Sharif:2015vya,Berezin:2016ubu,Babichev:2012sg,Misyura:2024fho,Khlopov:1999ys,Dymnikova:2015yma,Joshi:2013xoa,Joshi:2008zz,Joshi:1997de,Cai:2008mh,Simpson:2019cer,Mann:2021mnc,Culetu:2022otf,Baccetti:2018qrp,Vertogradov:2018ora,Vertogradov:2022zuo,Heydarzade:2023gmd,Vertogradov:2023uav}.

 \textcolor{black}{Oppenheimer and Snyder present the first rigorous model of gravitational collapse showing how a pressureless dust cloud can undergo continual contraction to form a black hole under general relativity \cite{Oppenheimer:1939ue}. Singh and Joshi analyze the end-state of spherical inhomogeneous dust collapse, demonstrating conditions under which naked singularities or black holes form based on initial density and velocity profiles \cite{Singh:1994tb}. Joshi and Singh identify a phase transition during the gravitational collapse of inhomogeneous dust, showing how variations in initial density profiles can lead to qualitatively different end-states such as black holes or naked singularities \cite{Joshi:1994br}. Casadio and Venturi analyze the semiclassical collapse of a homogeneous dust sphere, deriving the back-reaction of quantum effects on the dynamics and highlighting conditions for horizon formation and singularity avoidance \cite{Casadio:1995qy}. Jhingan et al. extend the analysis of spherical inhomogeneous dust collapse by examining how different initial data influence the causal structure of the resulting singularity, delineating criteria for black hole versus naked singularity outcomes \cite{Jhingan:1996jb}. Alberghi, Casadio, Vacca, and Venturi quantize the dynamics of a collapsing thin shell of matter, deriving its quantum wavefunction evolution and exploring implications for horizon formation and singularity avoidance \cite{Alberghi:1998xe}. Marković and Shapiro examine how a nonzero cosmological constant affects the dynamics of gravitational collapse, demonstrating its impact on collapse timescales and the resulting horizon structure \cite{Markovic:1999di}. Joshi et al. examine the physical and geometric factors leading to naked singularity formation in gravitational collapse, elucidating why and under what conditions cosmic censorship may fail \cite{Joshi:2001xi}. Gonçalves and Jhingan explore the role of radial pressure in gravitational collapse models, demonstrating how it affects singularity formation and the nature of the resulting spacetime singularities \cite{Goncalves:2001pf}.
 Harko examines the gravitational collapse of a Hagedorn fluid within the Vaidya geometry, deriving solutions that characterize horizon formation and singularity behavior in this context \cite{Harko:2003hs}. Joshi provides a comprehensive overview of the theoretical foundations and developments in gravitational collapse and spacetime singularities, examining both classical and quantum perspectives on singularity formation and visibility \cite{Joshi:2008zz}. Sharif and Kausar analyze gravitational collapse of a perfect fluid in $f(R)$ gravity, deriving the modified field equations and exploring conditions under which singularities form or are avoided \cite{Sharif:2010um}. Joshi and Malafarina review recent theoretical and numerical advances in gravitational collapse and spacetime singularities, emphasizing new insights into naked singularity formation and the status of the cosmic censorship conjecture \cite{Joshi:2011rlc}. Zhang, Zhu, Modesto, and Bambi investigate whether realistic gravitational collapse can produce static regular black holes and conclude that, under standard collapse dynamics, such static regular solutions cannot form without finely tuned matter profiles \cite{Zhang:2014bea}.
 Sharif and Waseem derive exact spherical dust solutions within the $f(R,T,R_{\mu\nu}T^{\mu\nu})$ gravity framework, illustrating how curvature–matter couplings modify collapse dynamics and energy conditions \cite{Sharif:2018sie}.  Mosani, Dey, and Joshi demonstrate that in nonmarginally bound dust collapse, strong curvature singularities can be globally visible under appropriate initial density and velocity profiles \cite{Mosani:2020mro}. Naidu, Bogadi, Kaisavelu, and Govender analyze the stability and horizon formation during dissipative gravitational collapse, highlighting how heat flux and anisotropic pressures influence the onset of horizon formation and the collapse end-state \cite{Naidu:2020oks}. 
Dey, Joshi, Mosani, and Vertogradov analyze the causal structure of singularities arising in non-spherical gravitational collapse, demonstrating how departures from spherical symmetry influence the formation and visibility of naked singularities \cite{Dey:2021jxw}. 
 Malafarina and Joshi edit a comprehensive volume that brings together leading research on the latest theoretical, numerical, and observational developments in gravitational collapse and spacetime singularities \cite{Malafarina:2024qdz}. Casadio et al. construct black hole models with a charged quantum dust core, examining how the core’s charge distribution modifies the spacetime geometry and horizon properties \cite{Casadio:2024lzd}.
Calmet, et al. explore the emergence of quantum hair during gravitational collapse, demonstrating that quantum effects can imprint nontrivial multipole moments on black hole spacetimes \cite{Calmet:2023met}. Casadio develops a model of black holes featuring quantum dust cores, illustrating how quantum corrections can replace singularities with extended core structures \cite{Casadio:2023ymt}. Gallerani et al. propose a framework for mass (re)distribution in quantum dust cores of black holes, exploring how quantum effects alter the interior mass profile and influence horizon properties \cite{Gallerani:2025wjc}.
 }

 A physically more appealing scenario is the formation of regular black holes through the gravitational collapse of astrophysical objects like massive stars \cite{Harko:2003hs}. Within this context, it remains unclear how the required exotic matter arises from ordinary baryonic matter. Recent proposals~\cite{Vertogradov:2024seh,Vertogradov:2025snh} suggest that a phase transition from dust or ordinary baryonic matter into a de Sitter-like state may naturally occur during stellar collapse, driven by conditions of extreme density and pressure. A study of ~\cite{Vertogradov:2025yto} investigated the gravitational collapse of dust transitioning to radiation, showing the possibility of negative pressure generation through an inhomogeneous phase transition rate, thus forming a regular black hole without singularities. {\color{black}Several works have investigated gravitational collapse, exotic matter fields, and modified gravity scenarios in the context of black hole and singularity formation. Barros, Dănilă, Harko, and Lobo construct and analyze black hole and naked singularity solutions sourced by three-form fields, detailing how the presence of these fields affects horizon structure and singularity properties \cite{Barros:2020ghz}.
Ziaie and Tavakoli investigate null fluid collapse within Rastall’s theory of gravity, demonstrating how the nonconservation of the energy–momentum tensor alters horizon formation and collapse end-states \cite{Ziaie:2019klz}.
 Creelman and Booth analyze the dynamics of null fluid collapse and demonstrate how specific conditions allow a collapsing null shell to undergo a bounce, avoiding singularity formation \cite{Creelman:2016laj}.
 Burikham et al. derive the theoretical minimum mass for spherically symmetric objects in arbitrary $D$-dimensional spacetimes and discuss its implications for addressing the mass hierarchy problem \cite{Burikham:2015nma}.
  Harko and Lake explore the dynamics of null fluid collapse in brane world scenarios, deriving solutions that illustrate how extra-dimensional effects influence horizon formation and energy conditions in these models \cite{Harko:2013sea}.
 Tavakoli, Marto, Ziaie, and Vargas Moniz study semiclassical gravitational collapse driven by a tachyon scalar field coupled to a barotropic fluid, revealing conditions under which horizons form or collapse is halted by quantum corrections \cite{Tavakoli:2013tpa}. Notably, Nicolini, Smailagic, and Spallucci introduce a noncommutative geometry–inspired Schwarzschild black hole model in which the mass is smeared over a minimal length scale, resulting in a regular, singularity-free geometry \cite{Nicolini:2005vd}. 
 Yavuz et al. derive exact solutions describing strange quark matter attached to a string cloud in spherically symmetric spacetimes admitting conformal motion and analyze their physical properties \cite{Yavuz:2005qb}.
  Kumar, Chatterjee, and Jaryal investigate the dynamics of gravitational collapse within pure Gauss-Bonnet gravity, revealing novel features distinct from standard general relativity \cite{Kumar:2025qqj}.
 Buoninfante, Di Filippo, Kolář, and Saueressig analyze dust collapse in quadratic gravity and identify the modified conditions under which horizons form in these higher-derivative theories \cite{Buoninfante:2024oyi}. }

However, realistic massive stars are not composed solely of dust; they are fundamentally baryonic in nature, transitioning through various phases including radiation and, critically, quark-gluon plasma. \textcolor{black}{Previous works, such as the analysis by Harko~\cite{Harko:2000ni}, indicate that purely quark matter collapse typically leads to naked singularities.} Hence, the interplay between baryonic and quark matter during collapse is crucial and requires careful examination. \textcolor{black}{Malafarina investigates the gravitational collapse of Hagedorn fluids and demonstrates that imposing a limiting Hagedorn temperature can halt collapse before singularity formation, potentially yielding stable equilibrium configurations rather than black holes \cite{Malafarina:2016yuf}.}

Given the astrophysical relevance and theoretical necessity to understand the formation of regular black holes from physically realistic initial conditions, the main aim of this work is to explore gravitational collapse models involving baryonic matter that undergoes a phase transition to quark-gluon plasma at high densities and temperatures. Specifically, we investigate how an inhomogeneous rate of matter transition impacts the spacetime geometry and pressure distributions. Our analysis demonstrates explicitly that collapse scenarios involving pure baryonic or pure quark matter inevitably lead to singularities, while allowing a controlled and inhomogeneous baryon-to-quark phase transition provides a robust mechanism for generating regular black hole solutions with a de Sitter-like core, thus avoiding the formation of singularities and satisfying observational constraints from recent EHT data.

This paper begins by outlining the motivation for studying regular black hole solutions and the fundamental limitations of classical general relativity in describing the interior structure of black holes. In particular, we highlight the problem of spacetime singularities, the role of energy conditions, and the possibility of resolving such issues through phase transitions into exotic states of matter—most notably, the emergence of a de Sitter core. In Section II, we focus on black hole solutions supported solely by quark matter. We examine how the inclusion of quark matter modifies the spacetime geometry and influences observable quantities such as the shadow radius of the black hole. A detailed phenomenological analysis is presented, investigating how variations in the quark matter contribution affect the shadow size, with comparisons to observational constraints from the Event Horizon Telescope. In Section III, the model is extended to include a composite system composed of both barotropic fluid and quark matter. This more realistic setup reflects the possibility that different phases of matter may coexist or dominate at various stages of gravitational collapse. Section IV explores the dynamical process of black hole formation under the assumption of a constant rate of energy exchange between the two fluid components. We analyze whether such interactions can give rise to regular, non-singular black hole solutions during the collapse. In Section V, we propose a more general and physically motivated model that allows for non-constant interaction rates between barotropic fluid and quark matter. We demonstrate that an inhomogeneous transition between these matter phases can dynamically lead to the formation of regular black holes. Subsection V.A presents an explicit analytical solution describing this collapse scenario, while Subsection V.B derives the mathematical and physical conditions necessary to ensure regularity at the core—specifically, the avoidance of curvature singularities and the realization of a de Sitter-like central structure. Finally, Section VI summarizes the main results of the paper, emphasizing the physical plausibility of forming regular black holes via realistic matter transitions, and discusses the broader implications for observational astrophysics.

\section{Exploring Black Hole Solutions Supported by Quark Matter}

We start our analysis by considering the most general dynamical, spherically symmetric spacetime expressed in Eddington-Finkelstein coordinates $\{v, r, \theta, \varphi\}$, given by the metric
\begin{equation}\label{eq:metric-general}
ds^2 = -\left(1-\frac{2M(v,r)}{r}\right)dv^2 + 2\varepsilon\, dv\,dr + r^2 d\Omega^2,
\end{equation}
where $M(v,r)$ represents the mass function depending explicitly on both the radial coordinate $r$ and advanced (Eddington-Finkelstein) time coordinate $v$. The parameter $\varepsilon=\pm 1$ signifies ingoing ($+1$) or outgoing ($-1$) energy flux, and $d\Omega^2 = d\theta^2 + \sin^2\theta\, d\varphi^2$ is the metric on the unit two-sphere. Without loss of generality, we choose $\varepsilon = +1$ to represent scenarios of gravitational collapse and black hole formation. The physical quantities associated with this spacetime geometry are given explicitly as follows:
\begin{align}\label{eq:physical-quantities}
\sigma(v,r) &= \frac{2\dot{M}(v,r)}{r^2},\\[2mm]
\rho(v,r) &= \frac{2M'(v,r)}{r^2},\\[2mm]
P(v,r) &= -\frac{M''(v,r)}{r},
\end{align}
where dots and primes denote partial derivatives with respect to $v$ and $r$, respectively. Here, $\sigma(v,r)$ corresponds physically to the density of energy flux, while $\rho(v,r)$ and $P(v,r)$ denote energy density and pressure of the collapsing matter, respectively. For strange quark matter, we adopt the MIT bag equation of state (EoS)~\cite{Farhi:1984qu,Witten:1984rs,Cheng:1998na}, which is commonly employed in astrophysical and cosmological contexts:
\begin{equation}\label{eq:eos-quark}
P = \frac{1}{3}\left(\rho - 4b\right),
\end{equation}
with $b$ denoting the bag constant. Physically, the parameter $b$ encapsulates the energy density difference between perturbative and true QCD vacua, thus quantifying the confinement energy necessary for quark matter stability. Imposing conservation of the energy-momentum tensor $T^{ik}_{\;\;\;;k}=0$ in this geometry yields:
\begin{equation}\label{eq:cons-law}
r\rho'(v,r) + 2[\rho(v,r) + P(v,r)] = 0.
\end{equation} Integrating Eq.~\eqref{eq:cons-law} with the EoS~\eqref{eq:eos-quark}, we find the quark matter density explicitly as:
\begin{equation}\label{eq:density-quark}
\rho(v,r) = b + C(v)\, r^{-8/3},
\end{equation}
where $C(v)$ is an integration function representing an effective radiation-like component arising naturally from the boundary conditions. From the definition of the density in Eq.~\eqref{eq:physical-quantities}, the corresponding mass function is:
\begin{equation}\label{eq:mass-quark}
M(v,r) = M_0(v) + \frac{b}{6} r^3 + \frac{3}{2} C(v)\, r^{1/3},
\end{equation}
with $M_0(v)$ representing the dynamical mass of the central object (black hole). Notice that the solution obtained here structurally resembles the Husain solution~\cite{Husain:1995bf,Vertogradov:2023uav}, incorporating a cosmological-like term ($\sim b\,r^3$) and a radiation-like component ($\sim C(v)\,r^{1/3}$). One can take $M_0$ and $C$ as a constant so the lapse function becomes 

\begin{equation} \label{sol1}
f(r)=1-\frac{2M_0}{r}-\frac{b}{3}r^2-3C\,r^{-2/3}.
\end{equation}
which emerges in a composite matter framework combining a barotropic fluid with quark matter. 

\begin{figure}[htp]
   \centering
\includegraphics[scale=0.8]{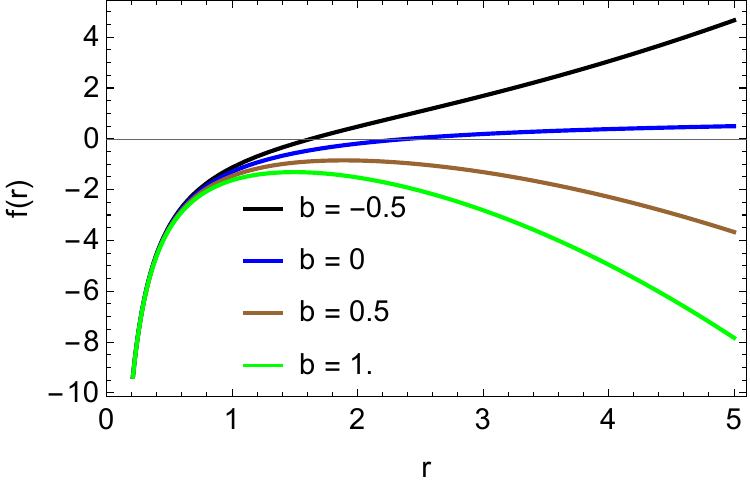}
\includegraphics[scale=0.8]{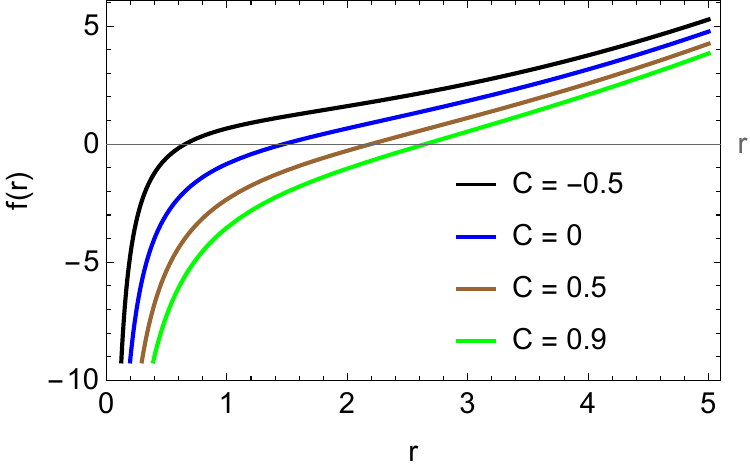}
    \caption{Figure shows lapse function $f(r)$ versus $r$ for different values of parameter $b$ and $C$. \textcolor{black}{ In the first panel of the figure, we take $M_0=1$ and $C=0.1$, on the other hand, in the second panel of the figure we take $M_0=1$  and $b=-0.5$.}}
    \label{fig:horizons1}
\end{figure}

\begin{figure}[htp]
   \centering
\includegraphics[scale=0.8]{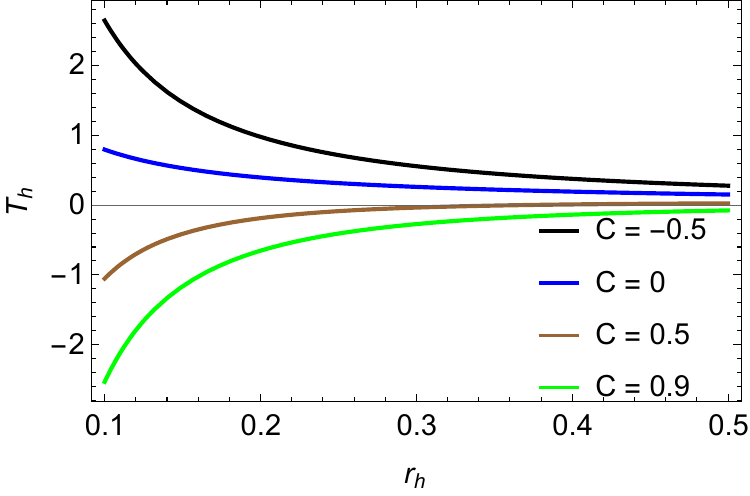}
\includegraphics[scale=0.8]{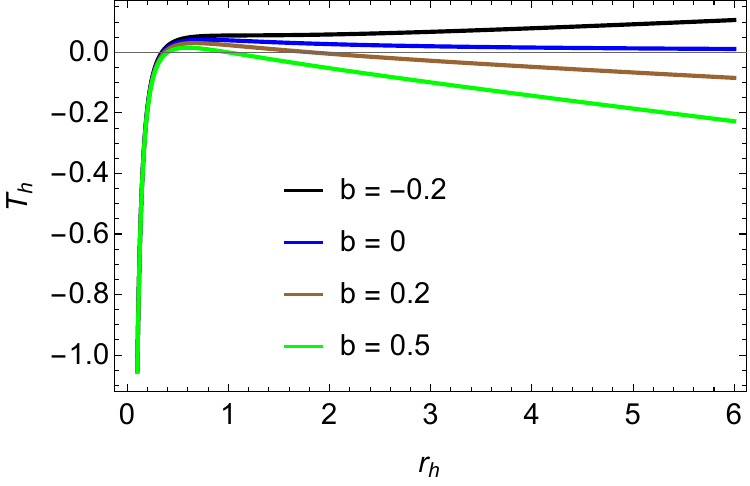}
    \caption{\textcolor{black}{Figure shows Hawking temperature $T_H$ versus $r_h$ for different values of parameter $b$ and $C$. \textcolor{black}{ In the first panel of the figure, we take $b=0.2$, on the other hand, in the second panel of the figure we take $C=0.5$.}}}
    \label{fig:temperature1}
\end{figure}

The standard Schwarzschild term, \(-2M_0/r\), encapsulates the mass contribution, while the \(-\frac{b}{3}r^2\) term plays the role of an effective cosmological constant or effect of the bag constant, governing the asymptotic behavior of the spacetime. Notably, the additional \(-3C\,r^{-2/3}\) term, arising from quark matter effects and a non-constant interaction rate with the barotropic fluid, modifies the gravitational potential in a nontrivial manner. This term's slower decay relative to the Schwarzschild term suggests a persistent influence on the spacetime geometry over a broad range of scales, potentially leading to alterations in the horizon structure as shown in Fig. \ref{fig:horizons1}. The interplay between these contributions affects both the location of the event horizon and the character of the central singularity, which may be softened relative to the classical Schwarzschild divergence. Furthermore, the modified metric function impacts thermodynamic quantities such as the surface gravity and entropy, thereby providing novel insights into black hole thermodynamics within the context of exotic matter sources. To assess whether the metric is singular at the origin, we evaluate its key curvature invariants in the limit as \(r \to 0\).
Ricci Scalar is calculated as
\begin{equation}
    R(r)=4b+\frac{4C}{3}\,r^{-8/3}\,.
\end{equation}
   As \(r\to0\), if \(C\neq 0\) then \(R\to\infty\) (diverges). On the other hand, Kretschmann scalar is calculated as 
  { \color{black}
\begin{equation}
 K(r)= \frac{8 b^{2}}{3}+\frac{16 C b}{9 r^{\frac{8}{3}}}+\frac{320 C M_0}{3 r^{\frac{17}{3}}}+\frac{48 M_0^{2}}{r^{6}}+\frac{568 C^{2}}{9 r^{\frac{16}{3}}}\,.
\end{equation}}
   As \(r\to0\), \(K\to\infty\), the Kretschmann scalar diverges. In any event, Ricci Squared diverges as \(r\to0\). Thus, even though the \(f(r)\) above might describe a composite matter black hole solution, it possesses a curvature singularity at \(r=0\) because the invariants diverge in that limit. Indeed, for a general barotropic EoS $P = \alpha \rho$, one obtains Husain-like mass functions:
\begin{equation}\label{eq:husain-general}
M(v,r) = M_0(v) + D(v)\,r^{1-2\alpha},
\end{equation}
with an energy density:
\begin{equation}\label{eq:husain-density}
\rho(v,r) = 2(1-2\alpha) \frac{D(v)}{r^{2\alpha+2}}.
\end{equation}
The mass of the black hole can be calculated by taking $f(r_h)=0$:
{\color{black}\begin{equation}
    M_0=\frac{r_h}{2}\left[1-\frac{b}{3}r_h^2-3C\,r_h^{-2/3}\right].
\end{equation}}
Then we derive the Hawking temperature from the surface gravity at the event horizon \(r_h\) (where \(f(r_h)=0\)). The surface gravity \(\kappa\) is defined by $
\kappa = \frac{1}{2}\,f'(r_h),$ and the Hawking temperature is then given by $T_H = \frac{\kappa}{2\pi} = \frac{1}{4\pi}\,f'(r_h),$ \textcolor{black}{yields}
{\color{black}
\begin{equation}
T_H =  \frac{1}{4\pi}\left[\frac{1}{r_h} - b\,r_h - C\,r_h^{-5/3} \right].
\end{equation}}
In the Fig. \ref{fig:temperature1}, varying \(C\) while keeping \(b\) fixed shows how the exotic quark matter contribution through the term affects the temperature, especially at smaller \(r_h\). In the second plot, the parameter \(b\) (associated with the effective cosmological or  the bag constant term) is varied. The \(b\) term influences the temperature more prominently at larger \(r_h\). These plots help illustrate the interplay between the different contributions to the Hawking temperature in this composite black hole solution.

\subsection{Effect of quark matter on the Shadow radius of the black hole}

\begin{figure}[htp]
   \centering
\includegraphics[scale=0.6]{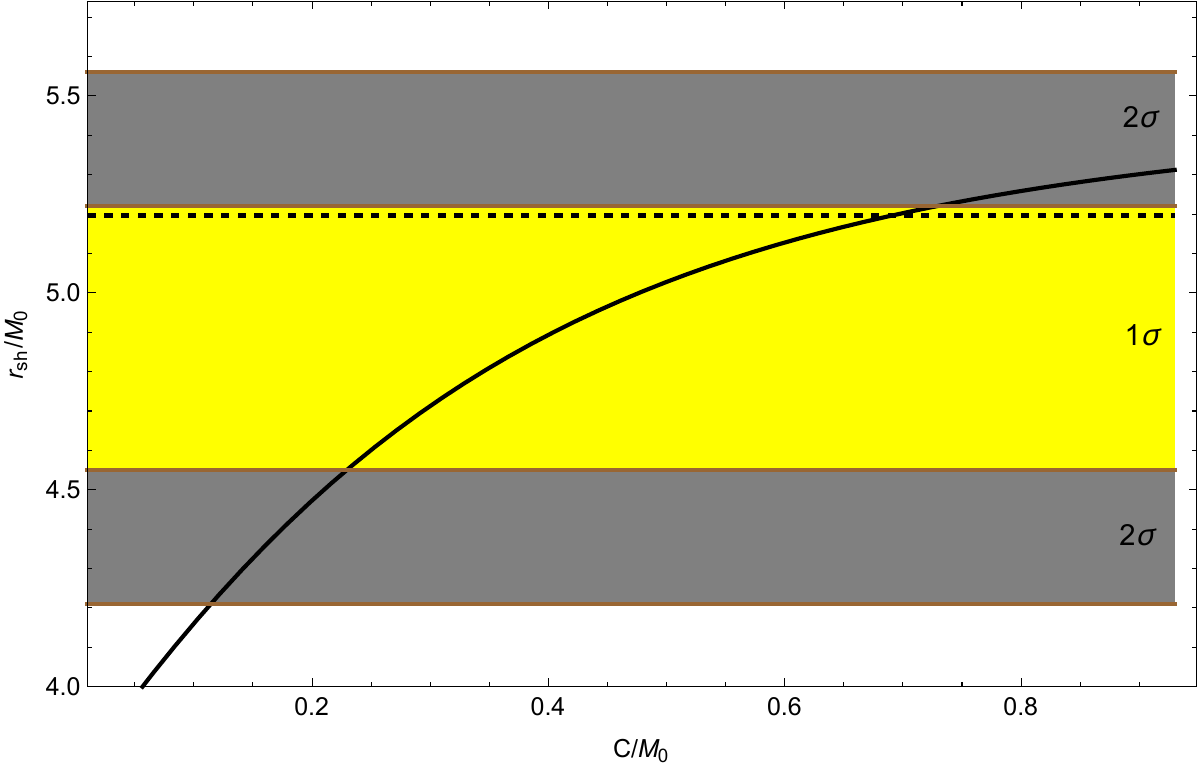}
\caption{Shadow radius $r_{sh}/M_0$ constraints of black hole supported by quark matter from the Event Horizon Telescope horizon-scale image of Sagittarius A* at $1\sigma$, after averaging the Keck and VLTI mass-to-distance ratio priors for the same with $M_0=1$ and \(\,b=-0.1\), and varying $C/M_0$. The dashed line denotes the Schwarzschild black hole’s shadow radius.}
    \label{fig:shadow1}
\end{figure}

Various recent studies have analyzed black hole shadows, gravitational lensing, and related phenomenology to explore modified gravity theories, quantum effects, and astrophysical observations of compact objects \cite{Falcke:1999pj,Bambi:2019tjh,Vagnozzi:2019apd,Allahyari:2019jqz,Pantig:2024asu,Lambiase:2024uzy,Lambiase:2024lvo,Yang:2023tip,Yang:2023agi,Luo:2023ndw,Pantig:2022gih,Zhang:2019glo,Lambiase:2023hng,Okyay:2021nnh,Ovgun:2018tua,Uniyal:2022vdu,Mustafa:2022xod,Kuang:2022xjp,Atamurotov:2022knb,Atamurotov:2013sca,Papnoi:2014aaa,Atamurotov:2015nra,Belhaj:2020rdb,Belhaj:2020okh,Meng:2022kjs,Ling:2021vgk,AraujoFilho:2024rss,Filho:2024zxx,Filho:2023ycx,Perlick:2015vta}.
For a photon propagating in the spacetime described by the metric \ref{sol1}, \textcolor{black}{the Lagrangian is
$2\mathcal{L} = g^{\mu\nu}p_{\mu}p_{\nu}=0,$}
with the four-momentum \(p_{\mu} = \frac{dx_{\mu}}{d\lambda}\) and \(\lambda\) the affine parameter. Assuming the photon moves in the equatorial plane (\(\theta=\pi/2\) and \(\dot{\theta}=0\)), the Hamiltonian constraint becomes
\begin{eqnarray}
-\frac{p_t^2}{2f(r)} + \frac{f(r)p_r^2}{2} + \frac{p_\phi^2}{2r^2}=0.
\end{eqnarray}
Because the metric is independent of \(t\) and \(\phi\), the corresponding Killing vectors yield the conserved quantities
$p_t=-E,\quad p_\phi=L,$ where \(E\) is the energy and \(L\) the angular momentum of the photon. The equations of motion then read
\begin{eqnarray}
\dot{t}=-\frac{p_t}{f(r)}=\frac{E}{f(r)},\quad \dot{\phi}=\frac{p_\phi}{r^2}=\frac{L}{r^2},\quad \dot{r}=f(r)p_r.
\end{eqnarray}
Defining the effective potential via
$\dot{r}^2+V_{\rm eff}(r)=0,$
and using the Hamiltonian constraint, we obtain
\begin{eqnarray}
V_{\rm eff}(r)=f(r)\Biggl(\frac{L^2}{r^2}-\frac{E^2}{f(r)}\Biggr).
\end{eqnarray}
For a circular photon orbit at \(r=r_p\), \cite{Claudel:2000yi} the conditions $V_{\rm eff}(r_p)=0$ and $\frac{dV_{\rm eff}}{dr}\Big|_{r=r_p}=0$ lead to the impact parameter
\begin{eqnarray}
\mu_p\equiv\frac{L}{E}=\left.\frac{r}{\sqrt{f(r)}}\right|_{r=r_p}.
\end{eqnarray}
The orbital equation follows from
\begin{eqnarray}
\frac{dr}{d\phi}=\frac{\dot{r}}{\dot{\phi}}=\frac{r^2f(r)p_r}{L},
\end{eqnarray}
which can be recast as
\begin{eqnarray}
\frac{dr}{d\phi}=\pm r\sqrt{f(r)\Biggl(\frac{r^2E^2}{f(r)L^2}-1\Biggr)}.
\end{eqnarray}
At the turning point \(r=r_p\) (where \(\frac{dr}{d\phi}=0\)), we have
\begin{eqnarray}
\frac{E^2}{L^2}=\frac{f(r_p)}{r_p^2},
\end{eqnarray}
so that the orbit equation becomes
\begin{eqnarray}
\frac{dr}{d\phi}=\pm r\sqrt{f(r)\Biggl(\frac{r^2f(r_p)}{f(r)r_p^2}-1\Biggr)}.
\end{eqnarray}
Now, consider the black hole shadow as seen by a static observer at \(r=r_o\). For a light ray leaving the observer at an angle \(\psi\) from the radial direction, one finds
\begin{eqnarray}
\cot\psi=\frac{\sqrt{g_{rr}}}{\sqrt{g_{\phi\phi}}}\,\frac{dr}{d\phi}\Bigg|_{r=r_o}
=\frac{1}{r_o\sqrt{f(r_o)}}\,\frac{dr}{d\phi}\Bigg|_{r=r_o}.
\end{eqnarray}
Substituting the orbit equation yields
\begin{eqnarray}
\cot^2\psi=\frac{r_o^2f(r_p)}{f(r_o)r_p^2}-1\quad\Longleftrightarrow\quad \sin^2\psi=\frac{f(r_o)r_p^2}{r_o^2f(r_p)}.
\end{eqnarray}
For \(r_p\), the shadow radius \(r_{sh}\) is
\begin{eqnarray}
r_{sh}=r_o\sin\psi=r_p\sqrt{\frac{f(r_o)}{f(r_p)}}.
\end{eqnarray}
\textcolor{black}{In figure \ref{fig:shadow1},} the black hole shadow radius increases with \(C\) but at a progressively slower rate, reflecting a balance between the negative \(\,b=-0.1\) term and the quark-matter contribution encoded by \(C\). Initially, when \(C\) is small, even small increments lead to a relatively rapid growth in the shadow radius, indicating that the quark-matter term strongly influences the near-horizon geometry shown in Table \ref{tab:shadowradius_table1}. As \(C\) becomes larger, however, its incremental effect begins to taper off. Overall, this behavior underscores how a negative \(b\) (representing a mild cosmological or bag parameter) moderates but does not fully counteract the quark-matter term, causing a gradual saturation in the shadow radius as \(C\) increases.

\begin{table}[h!]
\centering
\begin{tabular}{|c|c|}
\hline
\textbf{\textcolor{black}{\(C/M_0\)}} & \textbf{\textcolor{black}{\(r_{sh}/M_0\)}} \\ \hline
0.100 & 4.16 \\ \hline
0.200 & 4.47 \\ \hline
0.300 & 4.71 \\ \hline
0.400 & 4.89 \\ \hline
0.500 & 5.03 \\ \hline
0.600 & 5.13 \\ \hline
0.700 & 5.20 \\ \hline
0.800 & 5.26 \\ \hline
0.900 & 5.30 \\ \hline
\end{tabular}
\caption{\textcolor{black}{Shadow radius \(r_{sh}/M_0\) of black hole supported by quark matter for \(\,b=-0.1\) and different values of \(C/M_0\).}}
\label{tab:shadowradius_table1}
\end{table}

The solid curve in the figure shows how the shadow radius \(r_{sh}\), depends on the model parameter $C$ in the given metric. Physically, this means that larger values of \(C\) yield a bigger apparent size of the black hole shadow.  The colored bands correspond to observational constraints from the Event Horizon Telescope (EHT) imaging of Sagittarius A\(^*\) (Sgr A\(^*\)) \cite{Vagnozzi:2022moj}, combined with priors on the mass-to-distance ratio from optical/infrared measurements (Keck and VLTI). The green band denotes the region of parameter space compatible with the observed EHT image at the \(1\sigma\) level, while the surrounding gray band extends the acceptable range out to \(2\sigma\).  Physically, these bounds imply that for the metric model under consideration, only certain values of the parameter can reproduce a black hole shadow size consistent with EHT’s direct image of Sgr A\(^*\). This consistency check provides a way to test modifications of Schwarzschild spacetimes and constrain potential deviations in the near-horizon geometry of supermassive black holes. \textcolor{black}{Note that uncertainties in the measured value of the mass $M_0$ may lead to degeneracies with the obtained value of $C$ where different sets of $(M_0, C)$ may produce the same shadow.}

\section{Black Hole Solutions in a Composite Model of Barotropic Fluid and Quark Matter}

We now extend the analysis to a scenario involving both barotropic fluid and quark matter, initially assuming no energy exchange between the two fluids. Hence, their energy–momentum tensors are individually conserved. The barotropic fluid satisfies:
\begin{equation}\label{eq:barotropic-eos}
P_b = \alpha\rho_b,\quad\text{with}\quad \alpha\in[0,1],\quad\alpha\neq\frac{1}{2}.
\end{equation}
Separately conserving barotropic and quark components ($T^{ik}_{q;b\,;k}=0$), we find:
\begin{align}
\rho_b(v,r)&=\rho_{0b}(v)\,r^{-2(1+\alpha)},\label{eq:rho-barotropic}\\[2mm]
\rho_q(v,r)&=b + C(v)\,r^{-8/3}\label{eq:rho-quark},
\end{align}
with the integration functions $\rho_{0b}(v)$ and $C(v)$ determined by \textcolor{black}{boundary
conditions between baryonic and quark matter}. Consequently, the total mass function, integrating Einstein's equations, reads explicitly:
\begin{equation}\label{eq:mass-full}
M(v,r)=M_0(v)+\frac{\rho_{0b}(v)}{2(1-2\alpha)}\,r^{1-2\alpha}+\frac{b}{6}\,r^3+\frac{3}{2}\,C(v)\,r^{1/3}.
\end{equation}

\begin{figure}[htp]
   \centering
\includegraphics[scale=0.8]{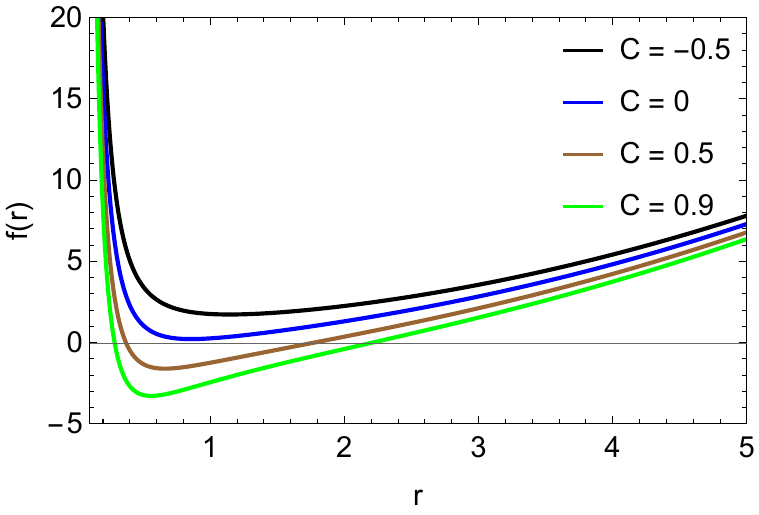}
\includegraphics[scale=0.8]{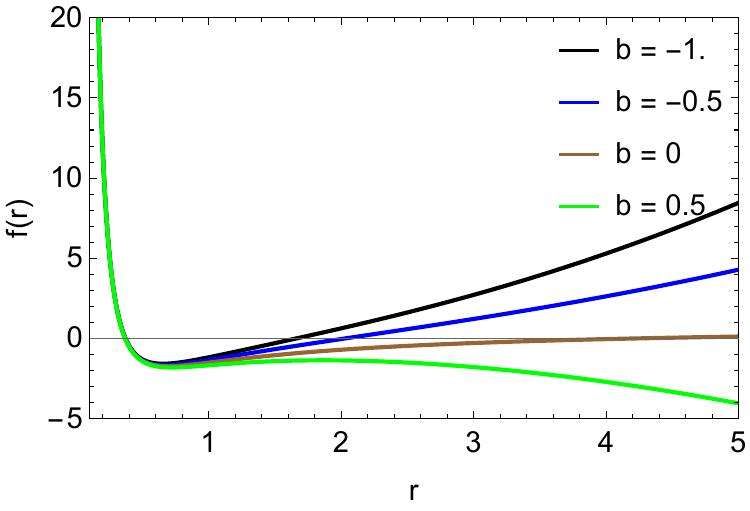}
    \caption{Figure shows lapse function $f(r)$ versus $r$ for different values of parameter $b$ and $C$. \textcolor{black}{In the first panel of the figure, we take $M_0=\alpha=\rho_{0b}=1$ and $b=-0.8$, on the other hand, in the second panel of the figure we take $C=0.5$ and $M_0=\alpha=\rho_{0b}=1$.}}
    \label{fig:horizons2}
\end{figure}

The term $M_0(v)$ represents the evolving black hole mass due to accretion. The barotropic component ($\rho_{0b}$-term) relates directly to the Husain solution. Energy conditions impose constraints on $\rho_{0b}(v)$, especially requiring negativity when $\alpha>1/2$, positivity for $\alpha<1/2$, and identifying it with the electric charge $-Q^2(v)$ when $\alpha=1$. The term proportional to $b\,r^3$ mirrors a cosmological constant-like contribution, producing non-asymptotic flatness.  The component proportional to $C(v)\,r^{1/3}$ behaves as radiation, scaling consistently with standard radiation-dominated universes. Remarkably, a straightforward non-interacting mixture of barotropic and quark matter inherently leads to singular solutions, as $M(v,r)$ does not vanish at the center ($r\to0$). Moreover, curvature invariants, such as the Kretschmann scalar, Ricci scalar, and Ricci-squared, explicitly diverge at the origin, confirming the inevitable central singularity. To physically obtain regular solutions at the origin, it becomes essential to include interactions between the quark and barotropic matter. This means allowing energy exchange and violating individual energy–momentum tensor conservation (while maintaining global conservation).

\begin{figure}[htp]
   \centering
\includegraphics[scale=0.8]{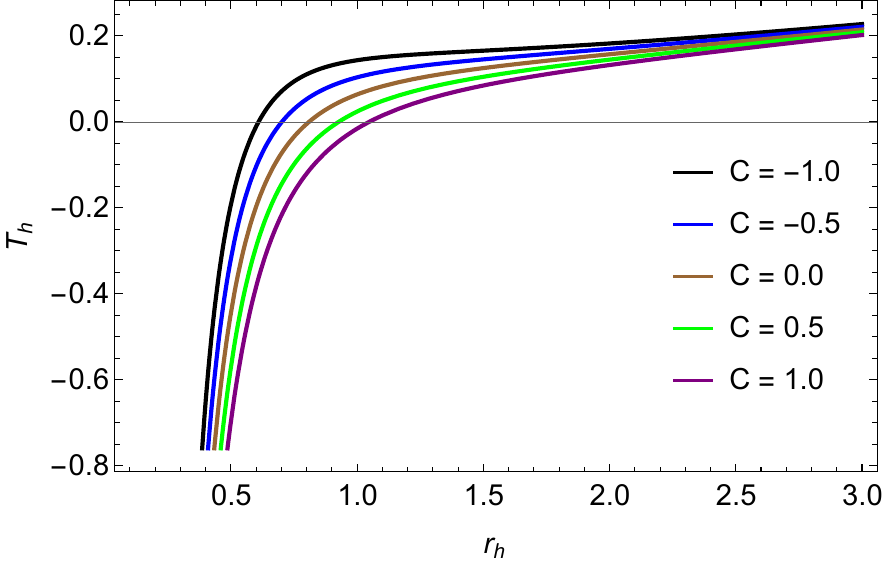}
\includegraphics[scale=0.8]{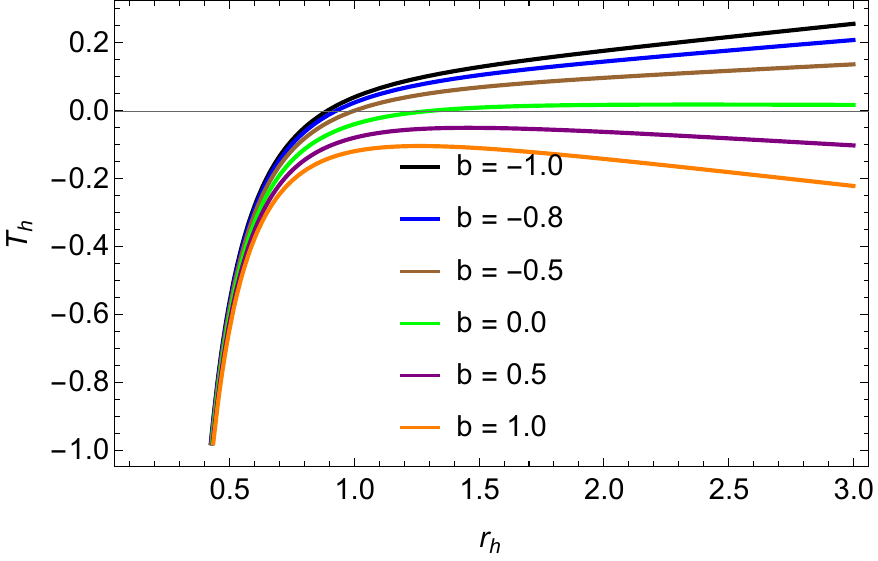}
    \caption{Figure shows Hawking temperature $T_H$ versus $r$ for different values of parameter $b$ and $C$. \textcolor{black}{In the first panel of the figure, we take $\alpha=\rho_{0b}=1$, and $b=-0.8$, on the other hand, in the second panel of the figure we take $C=0.5$ and $\alpha=\rho_{0b}=1$.}}
    \label{fig:temperature2}
\end{figure}

\textit{Effective equation of state (EoS):} We introduce an effective pressure and energy density, $P_{\text{eff}}$ and $\rho_{\text{eff}}$, respectively:
\begin{align}
P_{\text{eff}} &= \alpha\,\rho_{0b}(v)\,r^{-2(1+\alpha)} - b + \frac{C(v)}{3}\,r^{-8/3},\\[2mm]
\rho_{\text{eff}} &= \rho_{0b}(v)\,r^{-2(1+\alpha)} + b + C(v)\,r^{-8/3}.
\end{align}
The effective EoS parameter is thus $w_{\text{eff}}\equiv P_{\text{eff}}/\rho_{\text{eff}}$. \textcolor{black}{Figure \ref{fig:horizons2} plots the lapse function $f(r)$ versus $r$ by taking $M_0$ and $C$ as a constant. The roots of $f(r)$ (highlighted points) correspond to the locations of the horizons.} For the specific metric under consideration by taking $M_0$ and $C$ as a constant, the Kretschmann scalar takes the form

\begin{eqnarray}
K =\; &\left[-\frac{4M_0}{r^3}-\frac{2\alpha(2\alpha+1)\,\rho_{0b}}{1-2\alpha}\,r^{-2\alpha-2}-\frac{2b}{3}-\frac{10C}{3}\,r^{-\frac{8}{3}}\right]^2 \notag \\[1mm]
&+\frac{4}{r^2}\left[\frac{2M_0}{r^2}+\frac{2\alpha\,\rho_{0b}}{1-2\alpha}\,r^{-2\alpha-1}-\frac{2b}{3}\,r+2C\,r^{-\frac{5}{3}}\right]^2 \notag \\[1mm]
&+\frac{4}{r^4}\left[\frac{2M_0}{r}+\frac{\rho_{0b}}{1-2\alpha}\,r^{-2\alpha}+\frac{b}{3}\,r^2+3C\,r^{-\frac{2}{3}}\right]^2\,.
\end{eqnarray}
Substituting the explicit form of \(f(r)\) and its derivatives into Ricci scalar, we obtain
\begin{equation}
R = 2(1-\alpha)\,\rho_{0b}\,r^{-2\alpha-2}+4b+\frac{4\,C}{3}\,r^{-8/3}\,.
\end{equation}
\textcolor{black}{Notably, in the Schwarzschild limit where \(M_0\) is constant and all other parameters vanish (so that \(f=1-2M/r\)), the Ricci scalar correctly reduces to \(R=0\).} In contrast, the presence of additional \(r\)-dependent contributions from \(\rho_{0b}\) and \(C\) results in a nonvanishing Ricci scalar. For generic nonzero values of the parameters, the Kretschmann scalar diverges as \(r\to0\).  The Ricci scalar is given by
$    R = 2(1-\alpha)\,\rho_{0b}\,r^{-2\alpha-2}+4b+\frac{4C}{3}\,r^{-8/3}\,.$
   The terms with negative powers of \(r\) diverge in the limit \(r\to0\) (unless \(\rho_{0b}=0\) and \(C=0\)), while the constant term remains finite and does not offset the divergence. A similar analysis shows that
$R_{\mu\nu}R^{\mu\nu}\sim r^{-p}\quad\text{with } p>0\,,
$
   indicating a divergence as \(r\to0\). In summary, unless one fine-tunes the parameters (for example, by setting \(\rho_{0b}=0\) and \(C=0\)), all three invariants diverge as \(r\to0\). This divergence is indicative of a genuine curvature singularity at the origin. The mass of the black hole can be calculated by taking $f(r_h)=0$:
{\color{black}\begin{equation}
    M_0= \frac{r_h}{2}\left[ 1 - \frac{\rho_{0b}}{1-2\alpha}\,r_h^{-2\alpha} - \frac{b}{3}r_h^2 - 3C\,r_h^{-\frac{2}{3}}\right].
\end{equation}}
{\color{black} Then we calculate the Hawking temperature  as

\begin{equation}
T_H =\frac{1}{4\pi}\left[\frac{1}{r_h} - \rho_{0b}\,r_h^{-2\alpha-1} - b\,r_h - C\,r_h^{-\frac{5}{3}}\right].
\end{equation}}
Fig. \ref{fig:temperature2},  illustrates how the various contributions from the matter fields affect the thermal properties of the black hole. In the expression, the \(1/r_h\) term reflects the standard contribution from the Schwarzschild geometry, dominant at moderate radii. The term \(r_h^{-3}\) originates from the barotropic fluid component (with \(\rho_{0b}=1\) and \(\alpha=1\)); its inverse-cubic behavior implies a strong influence at very small \(r_h\), rapidly diminishing at larger radii. The linear \(-b\,r_h\) term, coming from the cosmological constant or extended bag parameter, becomes increasingly important at larger distances. Finally, the exotic quark matter contribution given by \(-C\,r_h^{-5/3}\) introduces a non-integer power-law behavior, which can affect both the near-horizon and asymptotic temperature profile depending on the value of \(C\). In the first plot (with fixed \(b=-0.8\)), varying \(C\) shows that higher values of \(C\) tend to lower the temperature-particularly noticeable at smaller radii-due to the enhanced negative contribution from the quark matter term. In the second plot (with fixed \(C=0.5\)), adjusting \(b\) shifts the temperature curve, where more negative values of \(b\) (which reduce the repulsive effect) lead to higher temperatures at larger \(r_h\), while more positive \(b\) accentuates a reduction in the temperature at larger distances. These behaviors highlight the sensitivity of the Hawking temperature to the interplay of standard gravitational effects and the corrections introduced by non-standard matter contributions.

\begin{figure}[htp]
   \centering
\includegraphics[scale=0.6]{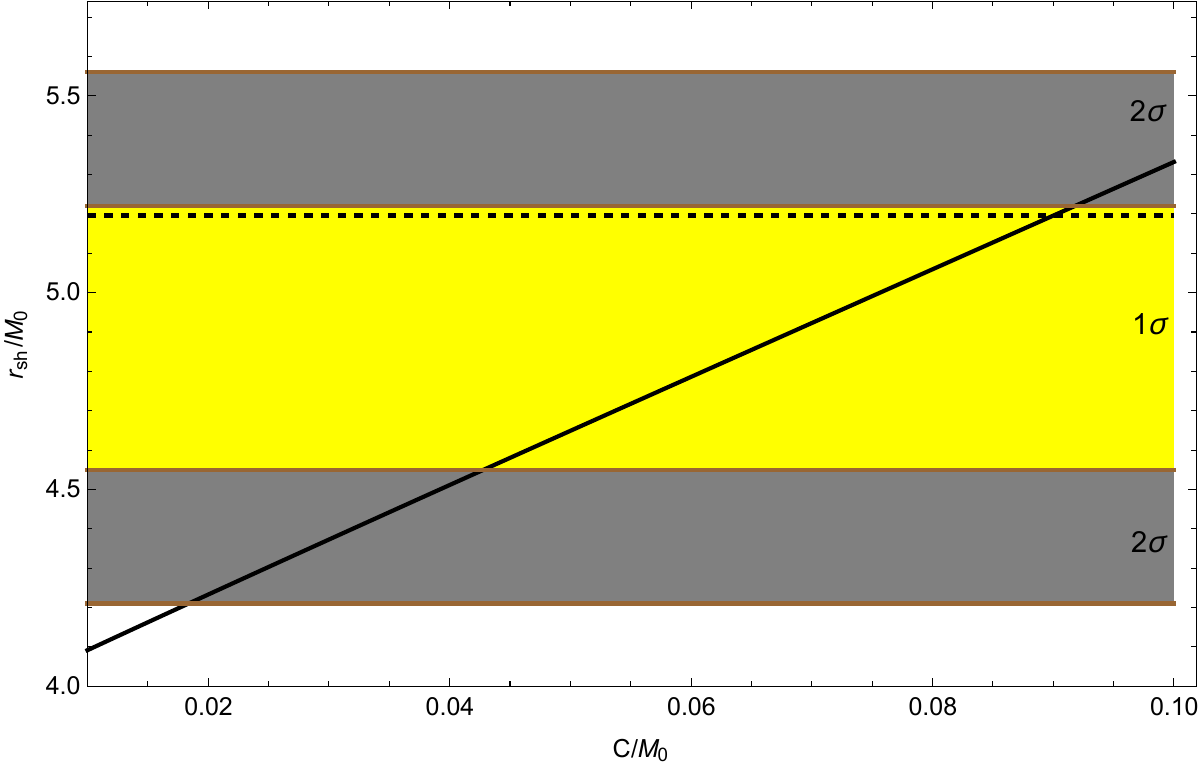}
\caption{\textcolor{black}{Constraints of shadow radius $r_{sh}/M_0$ of black hole supported by quark matter from the Event Horizon Telescope horizon-scale image of Sagittarius A* at $1\sigma$, after averaging the Keck and VLTI mass-to-distance ratio priors for the same with $M_0=1$, $b=-0.005$ and varying $C/M_0$. The dashed line denotes the Schwarzschild black hole’s shadow radius.}}
    \label{fig:shadow2}
\end{figure}

Similarly with previous section, we numerically calculate the shadow radius of the black hole in Table \ref{tab:shadowradius_table2} and in Fig. \ref{fig:shadow2}. In this plot, the black hole shadow radius  $r_{sh} $increases monotonically as the parameter \(C\) grows, with \(b\) fixed at \(-0.005\). Physically, a larger \(C\) strengthens the contribution from the exotic (quark-like) matter term in the spacetime geometry, which tends to expand the photon capture region and hence the apparent size of the black hole’s shadow. Although the parameter \(b\) introduces a mild cosmological or extended bag parameter effect (through a small negative value), the dominant influence in this figure clearly comes from \(C\), whose growth continuously increases the shadow radius for the parameter range shown. This behavior highlights the sensitivity of the near-horizon geometry—and thus the black hole shadow—to additional or unconventional matter components.

\begin{table}[h!]
\centering
\begin{tabular}{|c|c|}
\hline
\textbf{\(C/M_0\)} & \textbf{\(r_{sh}/M_0\)} \\ \hline
0.010 & 4.09 \\ \hline
0.020 & 4.23 \\ \hline
0.030 & 4.37 \\ \hline
0.040 & 4.51 \\ \hline
0.050 & 4.65 \\ \hline
0.060 & 4.79 \\ \hline
0.070 & 4.92 \\ \hline
0.080 & 5.06 \\ \hline
0.090 & 5.20 \\ \hline
\end{tabular}
\caption{\textcolor{black}{Shadow radius \(r_{sh}/M_0\) of black hole in a composite model of barotropic fluid and quark matter for $M_0=1$, $b=-0.005$ and different values of \(C/M_0\).}}
\label{tab:shadowradius_table2}
\end{table}

\section{Gravitational collapse with constant energy exchange between fluid components}

We now extend our analysis to a more physically realistic scenario, where the total energy-momentum tensor is conserved, yet interactions occur between the individual components of the collapsing matter, namely the barotropic fluid and the strange quark matter:
\begin{equation}
T^{ik}_{(\text{total});k} = 0\quad\Longrightarrow\quad T^{ik}_{(q);k}=-T^{ik}_{(b);k}.
\end{equation}
Physically, this condition means the barotropic matter is continuously converted into strange quark matter during gravitational collapse, at a constant dimensionless rate $\beta > 0$. The continuity equations for the individual components then take the modified form:
\begin{align}\label{eq:continuity-interacting}
r\rho_q' + 2(\rho_q + P_q) &= \beta\,\rho_q,\nonumber\\[2mm]
r\rho_b' + 2(\rho_b + P_b) &= -\beta\,\rho_q.
\end{align}
Substituting the equations of state into the continuity equations~\eqref{eq:continuity-interacting}, we obtain the explicit coupled system of equations:
\begin{align}\label{eq:coupled-density}
r\rho_q' + \left(\frac{8}{3}-\beta\right)\rho_q &= \frac{8}{3}b,\nonumber\\[2mm]
r\rho_b' + 2(1+\alpha)\rho_b &= -\beta\,\rho_q.
\end{align}
Solving the first equation explicitly yields the quark matter density:
\begin{equation}\label{eq:density-quark-interaction}
\rho_q(v,r)=\frac{b}{2}+C(v)\,r^{\,\beta-\frac{8}{3}}.
\end{equation}
We note immediately that when the interaction rate vanishes ($\beta=0$), Eq.~\eqref{eq:density-quark-interaction} reduces precisely to the non-interacting case. Using the solution~\eqref{eq:density-quark-interaction} in the second equation of~\eqref{eq:coupled-density}, we obtain after integration the density of barotropic matter undergoing continuous conversion into quark matter:
\begin{equation}\label{eq:density-barotropic-interaction}
\rho_b(v,r)=-\frac{\beta b}{4(1+\alpha)}-\frac{\beta\, C(v)}{\beta+2\alpha-\frac{2}{3}}\,r^{\,\beta-\frac{8}{3}}+E(v)\,r^{-2(1+\alpha)}.
\end{equation}
Here, $E(v)$ is an integration function determined by initial and boundary conditions of the collapse. Consequently, the \emph{effective energy density} $\rho_{\text{eff}}=\rho_q+\rho_b$ becomes:
{\color{black}\begin{equation}\label{eq:effective-density-interaction}
\rho_{\text{eff}}(v,r)=\xi+\eta(v)\,r^{\,\beta-\frac{8}{3}}+E(v)\,r^{-2(1+\alpha)},
\end{equation}}
where for compactness we define the constants:
\begin{align}\label{eq:xi-eta}
\xi&=\frac{b}{2}\left(1-\frac{\beta}{2+2\alpha}\right),\\[2mm]
\eta(v)&=C(v)\left(1-\frac{\beta}{\beta+2\alpha-\frac{2}{3}}\right).
\end{align}
In the non-interacting limit ($\beta\to0$), these constants revert to the previous results:
\begin{align}
\lim_{\beta\to0}\xi&=\frac{b}{2},\\[2mm]
\lim_{\beta\to0}\eta(v)&=C(v).
\end{align}
From Einstein’s equations, the effective energy density relates directly to the mass function by:
\begin{equation}\label{eq:Einstein-mass-interaction}
\rho_{\text{eff}}=\frac{2M'}{r^2},
\end{equation}
allowing explicit integration to yield the general form of the mass function as:
\begin{equation}\label{eq:mass-function-interaction}
M(v,r)=M_0(v)+\frac{\xi}{6}\,r^3+\frac{\eta(v)}{2\left(\beta+\frac{1}{3}\right)}\,r^{\,\beta+\frac{1}{3}}+\frac{E(v)}{2(1-2\alpha)}\,r^{\,1-2\alpha}.
\end{equation}
\textcolor{black}{The structure of Eq.~\eqref{eq:mass-function-interaction} clearly shows the consistency of our model, reducing exactly to the non-interacting scenario when $\beta=0$. Noticeably, the solution \eqref{eq:mass-function-interaction} with \eqref{eq:metric-general} is not asymptotically flat due to the presence of a cosmological-constant-like term $\sim r^2$, similar to previous known solutions modeling matter conversion scenarios~\cite{Vertogradov:2025snh}.} An important physical inquiry arises: can a continuous conversion of barotropic matter into strange quark matter, at a constant dimensionless interaction rate $\beta$, yield a regular black hole solution (free of central singularities)? Careful analysis shows that, similarly to the non-interacting scenario, the mass function~\eqref{eq:mass-function-interaction} does not vanish as $r\to 0$, unless special fine-tuning conditions are imposed. Thus, the spacetime generically exhibits a curvature singularity at the origin. Indeed, explicit calculation of curvature invariants, such as the Kretschmann scalar ($K$), Ricci scalar ($R$), and Ricci squared ($R_{\mu\nu}R^{\mu\nu}$), reveals divergences as $r\to0$. The effective equation of state (EoS) describing the matter supporting this spacetime is characterized by:
\begin{align}
P_{\text{eff}}(v,r)&=-\xi-\frac{\eta(v)}{2}\left(\beta-\frac{2}{3}\right)r^{\,\beta-\frac{8}{3}}+\alpha E(v)r^{-2(1+\alpha)},\\[2mm]
w_{\text{eff}}&=\frac{P_{\text{eff}}}{\rho_{\text{eff}}},
\end{align}
clearly indicating exotic matter behavior near the center ($r\to0$), where both density and pressure become singular unless carefully tuned. Our analysis clearly indicates that gravitational collapse scenarios with constant energy exchange between barotropic and quark matter generally result in singular black holes. All curvature invariants: Kretschmann scalar, Ricci scalar, Ricci squared similarly diverge at the center unless special fine-tuning conditions eliminate these terms. Such divergences reflect genuine physical curvature singularities, emphasizing that interaction alone—at least with a constant rate—does not naturally yield regular black hole solutions. More complex interaction dynamics are thus needed to achieve regularity at the center.

\section{Regular Black Hole Solutions Arising from Non-Constant Interactions between Barotropic Fluid and Quark Matter}

We now consider a more realistic scenario in which barotropic matter continuously transforms into strange quark matter during gravitational collapse, with a dimensionless interaction rate $\beta(v,r)$ that is no longer constant. Specifically, we assume that $\beta(v,r)$ intensifies as the center ($r\to0$) is approached, satisfying:
\begin{equation}\label{eq:beta-condition}
\beta=\beta(v,r),\quad\text{with}\quad \frac{\partial \beta}{\partial r}<0.
\end{equation}
Such spatial dependence allows greater flexibility in achieving regular solutions at the origin. Incorporating this into the continuity equations, we have explicitly:
\begin{align}\label{eq:system-nonconstant-beta}
r\rho_q'+\left(\frac{8}{3}-\beta(v,r)\right)\rho_q&=\frac{8}{3}b,\\
r\rho_b'+2(1+\alpha)\rho_b&=-\beta(v,r)\rho_q.
\end{align}
In general, without specifying the explicit functional form of $\beta(v,r)$, these equations cannot be integrated analytically. Nevertheless, formal integral solutions can be expressed. For the quark matter density, we have:
\begin{equation}\label{eq:formal-quark-density}
\rho_q(v,r)=C(v)e^{\int\frac{\beta-\frac{8}{3}}{r}dr}+\frac{8}{3}b\,e^{\int\frac{\beta-\frac{8}{3}}{r}dr}\int\frac{dr}{r\,e^{\int\frac{\beta-\frac{8}{3}}{r}dr}},
\end{equation}
while the barotropic matter density takes the form:
\begin{equation}\label{eq:formal-barotropic-density}
\rho_b(v,r)=r^{-2(1+\alpha)}\left(E(v)-\int\beta(v,r)\,r^{1+2\alpha}\rho_q(v,r)\,dr\right).
\end{equation}
The effective energy density is thus formally expressed as:
\begin{equation}\label{eq:effective-density-general}
\rho_{\text{eff}}(v,r)=\rho_q(v,r)+\rho_b(v,r),
\end{equation}
and the corresponding mass function is obtained by integrating Einstein’s equations as:
\begin{equation}\label{eq:mass-function-general}
M(v,r)=M_0(v)+\frac{1}{2}\int\rho_{\text{eff}}(v,r)\,r^2\,dr.
\end{equation}

\subsection{Explicit model leading to a regular black hole}

To concretely illustrate how a suitable choice of the function $\beta(v,r)$ can lead to a regular black hole solution, we now specify:
\begin{equation}\label{eq:beta-explicit}
\beta(v,r)=\frac{5}{3}-a(v)\,r,\quad a(v)>0.
\end{equation}
This choice, slightly different from that proposed in~\cite{Vertogradov:2025snh}, facilitates analytical integration. Solving the equations explicitly, we find the quark matter density to be:
\begin{equation}\label{eq:rho-q-explicit}
\rho_q(v,r)=\frac{c(v)}{r}e^{-a(v)r}+\frac{8b}{3a(v)r}.
\end{equation}
For general values of $\alpha$, integration for barotropic density is complicated. Hence, we specialize initially to dust collapse ($\alpha=0$), obtaining explicitly:
\begin{equation}\label{eq:rho-b-explicit}
\rho_b(v,r)=\frac{1}{r^2}\left[e(v)+\frac{2c(v)}{3a(v)}e^{-a(v)r}-\frac{40b\,r}{9a(v)}-r\,c(v)e^{-a(v)r}+\frac{4b\,r^2}{3}\right].
\end{equation}
Summing \eqref{eq:rho-q-explicit} and \eqref{eq:rho-b-explicit}, the effective energy density simplifies to:
\begin{equation}\label{eq:rho-eff-explicit}
\rho_{\text{eff}}(v,r)=\frac{4b}{3}+\frac{3a(v)e(v)+2c(v)e^{-a(v)r}-\frac{16}{3}b\,r}{3a(v)\,r^2}.
\end{equation}
Integrating Eq.~\eqref{eq:rho-eff-explicit}, we obtain the explicit mass function:
\begin{equation}\label{eq:mass-explicit}
M(v,r)=M_0(v)-\frac{c(v)}{3a(v)^2}e^{-a(v)r}+\frac{e(v)}{2}r-\frac{4b}{9a(v)}r^2+\frac{2b}{9}r^3.
\end{equation}

\subsection{Conditions for regularity at the center}

We now determine the conditions under which the solution \eqref{eq:mass-explicit} describes a regular black hole at the origin ($r=0$). 

\textbf{Regularity demands two essential criteria}:

1) To avoid singularities at the center, $\rho_{\text{eff}}$ in Eq.~\eqref{eq:rho-eff-explicit} must be finite as $r\to0$, imposing the condition:
\begin{equation}\label{eq:condition-a}
a(v)=-\frac{2\cdot 3^{1/3}}{3}\frac{b^{1/3}}{M_0(v)^{1/3}}.
\end{equation}

2) To ensure no conical singularities and that spacetime is regular at $r=0$, the mass function must vanish, $M(v,0)=0$. This condition yields explicit constraints on integration functions:
\begin{align}\label{eq:condition-c-e}
c(v)&=\frac{4\,b^{2/3}\,M_0(v)^{1/3}}{3^{1/3}},\\[2mm]
e(v)&=\frac{4\,b^{1/3}\,M_0(v)^{2/3}}{3^{2/3}}.\label{eq:condition-e}
\end{align}
Crucially, positivity of the mass $M_0(v)>0$ demands the bag constant to be negative ($b<0$). Although unusual, this scenario physically corresponds to a modified vacuum structure of quark matter supporting regular central cores. Under conditions \eqref{eq:condition-a}, \eqref{eq:condition-c-e}, and \eqref{eq:condition-e}, the mass function explicitly becomes:
\begin{equation}\label{eq:mass-final-regular}
M(v,r)=M_0(v)-\frac{c(v)}{3a(v)^2}e^{-a(v)r}+\frac{e(v)}{2}r-\frac{4b}{9a(v)}r^2+\frac{2b}{9}r^3,
\end{equation}
fully regular at $r=0$. Evaluating curvature invariants explicitly at the center confirms regularity by taking $M_0$, $c$, $a$ and $e$ as a constant. Specifically, as $r\to0$, one finds finite limits for:

Kretschmann scalar:
\begin{equation}
\lim_{r\to0}K(r)=16\,\kappa^2,
\end{equation}

Ricci scalar:
\begin{equation}
\lim_{r\to0}R(r)=4\,\kappa,
\end{equation}

Ricci squared invariant:
\begin{equation}
\lim_{r\to0}R_{\mu\nu}R^{\mu\nu}(r)=4\,\kappa^2,
\end{equation}
where the finite constant $\kappa$ is defined explicitly as:
\begin{equation}
\kappa=\frac{a c}{18}+\frac{2b}{9}.
\end{equation}
All curvature invariants remain finite, confirming the absence of physical singularities at the center. This explicit model demonstrates a physically compelling scenario: gravitational collapse accompanied by a spatially increasing interaction rate can yield \textit{regular black hole solutions}. The spatial dependence naturally avoids singularities, forming nonsingular quark-matter cores within collapsing objects. Although the requirement $b<0$ implies exotic conditions, this scenario provides a valuable theoretical prototype highlighting the importance of interaction dynamics in gravitational collapse scenarios and singularity avoidance. \textcolor{black}{It is worth mentioning a few words about the solution \eqref{eq:mass-final-regular}: evidently, the phase transition does not occur instantaneously, but rather at a specific radius where critical matter densities are reached. Therefore, this solution should be considered only as an approximation valid near the center of the collapsing matter and is not applicable for describing the exterior spacetime. Consequently, at a certain radius-determined not arbitrarily but precisely based on the properties of the initial matter and the new matter formed during the phase transition-the solution~\eqref{eq:mass-final-regular} must be matched to an external solution that describes the original matter content of the star.}
\textcolor{black}{Moreover, we note that the obtained solution can not be smoothly matched to the Vaidya metric, since the energy density and pressure do not vanish at physically reasonable distances from the center; instead, they take finite values. Because of this, a smooth matching requires the introduction of a new thin matter layer , which allows both the interior and exterior solutions to be smoothly joined~\cite{Poisson:2009pwt}. However, if we consider the Husain or Kiselev solution as the exterior one, such matching can still be performed in this model by satisfying just three conditions: at the phase transition radius, the mass function, energy density, and pressure of both solutions must coincide.}
\textcolor{black}{It is important to note that the vanishing of the mass function at the center, $\lim_{r \to 0} M(v, r) = 0$, does not imply that the black hole mass $M_0(v)$ itself must be zero. If $M_0(v) \equiv 0$, then the core of the regular black hole would reduce to a Minkowski core rather than a de Sitter one. It should also be emphasized that, in this case, the energy flux does not exhibit any distinguishing features compared to well-known models such as those of Bardeen, Hayward, or Dymnikova, which similarly satisfy the condition $\lim_{r \to 0} M(v, r) = 0$ while maintaining a non-zero black hole mass $M_0(v) \neq 0$.}

\section{Conclusion}

Regular black holes have attracted substantial attention due to their promise of addressing the central singularity problem in classical black hole physics. Central to all regular black hole models is the violation of the strong energy condition near the core, thereby circumventing classical singularity theorems. Despite extensive theoretical research, the physical reality and formation mechanisms underlying regular black holes remain incompletely understood, largely due to the requisite exotic matter at their cores, which is not directly observed in astrophysical environments.

In an effort to provide a physically realistic scenario, previous studies have hypothesized that such exotic matter arises through gravitational collapse and subsequent phase transitions of ordinary baryonic matter at critical densities and temperatures. Building on earlier foundational work, such as the gravitational collapse model of dust transitioning into radiation discussed in~\cite{Vertogradov:2025snh}, we analyzed the collapse of baryonic matter undergoing a transition into quark matter. Our findings demonstrate explicitly that a constant dimensionless rate of transition inevitably leads to the formation of a singular black hole. Crucially, however, allowing for an inhomogeneous transition rate-accelerating as matter approaches the central region—enables the formation of a regular black hole with a nonsingular de Sitter-like core.

An essential unresolved aspect of this model pertains to determining the precise functional form of the dimensionless transition parameter, \( \beta \), which is expected to emerge from quantum effects at high densities. While our model effectively captures the core region’s physics, it lacks explicit methods for observationally testing the presence or absence of a singularity directly. Indeed, our analysis of the black hole shadow radius demonstrates sensitivity to the presence of quark matter, offering promising observational signatures that can be constrained by Event Horizon Telescope (EHT) data.

Moreover, a distinctive observational signature of the proposed phase transition scenario could involve a significant energy release prior to apparent horizon formation, potentially observable as a transient astrophysical event. Future studies employing detailed numerical simulations, combined with ongoing and future observational efforts from facilities like the EHT, will be crucial in refining theoretical models and validating their predictions.

In summary, this work advances our theoretical understanding of regular black hole formation by highlighting the crucial role of non-constant matter transitions. Nevertheless, key challenges remain, including identifying microscopic mechanisms underlying these transitions, developing predictive observational signatures, and rigorously testing theoretical predictions against astrophysical observations.

\acknowledgments
V.V. gratefully acknowledges the hospitality extended by Eastern Mediterranean University. A.{\"O}. would like to acknowledge the contribution of the COST Action CA21106 - COSMIC WISPers in the Dark Universe: Theory, astrophysics and experiments (CosmicWISPers), the COST Action CA21136 - Addressing observational tensions in cosmology with systematics and fundamental physics (CosmoVerse), the COST Action CA22113 - Fundamental challenges in theoretical physics (THEORY-CHALLENGES), the COST Action CA23130 - Bridging high and low energies in search of quantum gravity (BridgeQG) and the COST Action CA23115 - Relativistic Quantum Information (RQI) funded by COST (European Cooperation in Science and
Technology). We also thank EMU, TUBITAK, ULAKBIM (Turkiye) and SCOAP3 (Switzerland) for their support.  

\bibliographystyle{apsrev4-1}
\bibliography{ref}

\end{document}